\documentclass[10pt,conference]{IEEEtran}

\usepackage[T1]{fontenc}
\usepackage[utf8]{inputenc}
\usepackage[american]{babel}
\usepackage{csquotes}
\usepackage{microtype}

\makeatletter
\MT@redefine@patch{eqnum}{%
  \MT@patch@patch\tagform@{(}{\leftprotrusion{(}}%
  \MT@patch@patch\tagform@{)}{\rightprotrusion{)}}%
  \MT@ifdefined@n@TF{eqref }
    {\MT@exp@cs\MT@patch@patch{eqref }}{\MT@patch@patch\eqref}
       {\tagform@}{\@nameuse{MT@patch@saved@\string\tagform@}}%
}{}
\makeatother

\usepackage{paralist}
\setdefaultenum{(1)}{(a)}{(i)}{A.}
\usepackage{enumitem}
\usepackage{url}
\usepackage{flushend}
\usepackage{etoolbox}
\usepackage{fix-cm}
\usepackage{textpos}

\usepackage{amsmath}
\usepackage{amssymb}
\usepackage{bm}
\usepackage{relsize}
\usepackage{siunitx}
\usepackage[super]{nth}

\usepackage{booktabs}
\usepackage{multirow}
\usepackage{colortbl}
\usepackage{makecell}
\usepackage{rotating}
\usepackage{threeparttable}

\usepackage{float}
\usepackage{graphicx}

\usepackage{listingsutf8}

\usepackage{xparse}
\usepackage{xspace}

\usepackage[xindy,acronym]{glossaries}

\usepackage[numbers,sort&compress]{natbib}

\usepackage[hidelinks,bookmarks=false]{hyperref}
\usepackage[all]{hypcap}

\usepackage[capitalise]{cleveref}

\begin{document}

\title{Challenges of Testing an Evolving Cancer Registration Support System in Practice}

\author{
\IEEEauthorblockN{Christoph~Laaber, Tao~Yue, Shaukat~Ali}
\IEEEauthorblockA{Simula Research Laboratory\\
Oslo, Norway\\
\{laaber, tao, shaukat\}@simula.no}
\and
\IEEEauthorblockN{Thomas~Schwitalla}
\IEEEauthorblockA{Cancer Registry of Norway\\
Oslo, Norway\\
thsc@kreftregisteret.no}
\and
\IEEEauthorblockN{Jan~F.~Nygård}
\IEEEauthorblockA{Cancer Registry of Norway\\
Oslo, Norway\\
UiT The Arctic University of Norway\\
Tromsø, Norway\\
jfn@kreftregisteret.no}
}

\maketitle

\begin{textblock*}{\textwidth}(-1.5em,18cm)
	\fbox{
		\begin{minipage}{\linewidth}
			\footnotesize \textcopyright{} 2023 IEEE. Personal use of this material is permitted. Permission from IEEE must be obtained for all other uses, in any current or future media, including reprinting/republishing this material for advertising or promotional purposes, creating new collective works, for resale or redistribution to servers or lists, or reuse of any copyrighted component of this work in other works.
			DOI: \href{https://doi.org/10.1109/ICSE-Companion58688.2023.00102}{10.1109/ICSE-Companion58688.2023.00102}
		\end{minipage}
	}
\end{textblock*}

\begin{abstract}

The Cancer Registry of Norway (CRN) is a public body responsible for capturing and curating cancer patient data histories to provide a unified access to research data and statistics for doctors, patients, and policymakers.
For this purpose, CRN develops and operates a complex, constantly-evolving, and socio-technical software system.
Recently, machine learning (ML) algorithms have been introduced into this system to augment the manual decisions made by humans with automated decision support from learned models.
To ensure that the system is correct and robust and cancer patients' data are properly handled and do not violate privacy concerns, automated testing solutions are being developed.
In this paper, we share the challenges that we identified when developing automated testing solutions at CRN.
Such testing potentially impacts the quality of cancer data for years to come, which is also used by the system's stakeholders to make critical decisions.
The challenges identified are not specific to CRN but are also valid in the context of other healthcare registries. We also provide some details on initial solutions that we are investigating to solve the identified challenges.

\end{abstract}


\section{Introduction}\label{introduction}
\begin{figure*}[ht]
\includegraphics[width=1.0\textwidth]{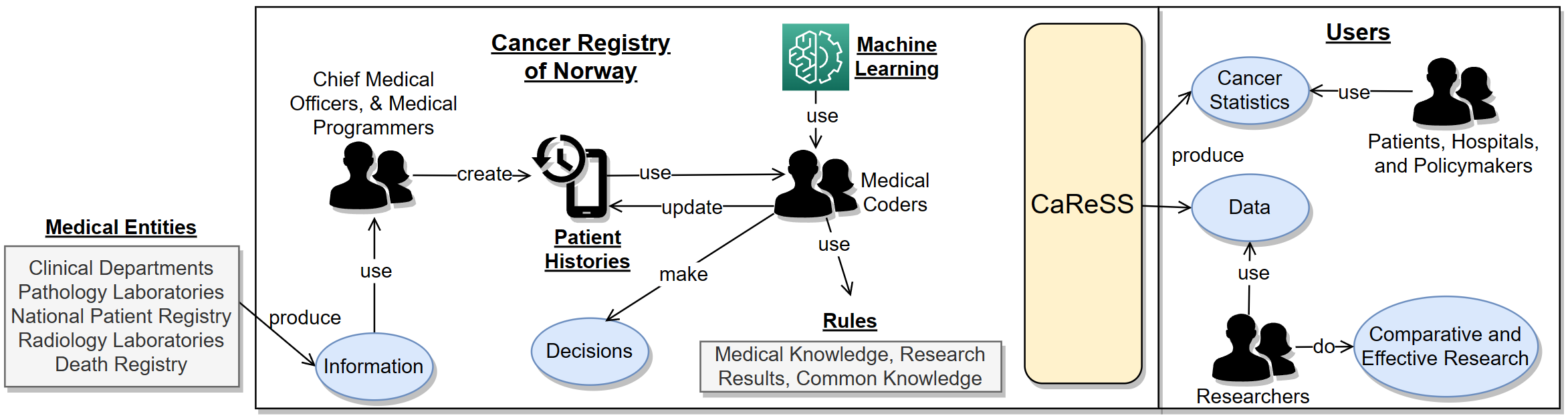}
\centering
\caption{Overview of the Cancer Registration Support System.}
\label{overview}
\vspace{-10pt}
\end{figure*}

The Cancer Registry of Norway (CRN) is a public entity responsible for collecting cancer patients' data from multiple sources, including diagnostic, treatment, and follow-up data, as shown in Figure~\ref{overview}. Next, such data are processed for cancer research and statistics to various users such as patients, medical institutions, and relevant governmental authorities.
For this purpose, CRN has developed a Cancer Registration Support System (CaReSS), an interactive, human-in-the-loop decision support system.
CaReSS experiments with using machine learning (ML) techniques and is subject to continuous evolution due to, e.g., new diagnostic methods and treatment, standards, and regulations (e.g., General Data Protection Regulation (GDPR)), in addition to bug fixing and updates to its ML models.

CaReSS collects patient information from various medical entities such as pathology and radiology laboratories, hospitals, and other registries such as the national patient registry and the death registry, as shown in Figure~\ref{overview}.
Such patient information is recorded as \textit{cancer messages} and prepared for trained \textit{medical coders} to create \textit{cancer cases}, each of which is a timeline of a patient's diagnostic tests, treatments and follow-ups.
The coding process, i.e., transforming medical reports (messages) into uniform codes based on a classification system such as the International Classification of Diseases
(ICD)\footnote{\url{https://www.who.int/standards/classifications/classification-of-diseases}} from World Health Organization (WHO), is dependent on hundreds of \textit{medical rules}, which are used to validate cancer messages, assigning these to cancer cases, etc. These rules are defined based on the domain knowledge of medical experts. The CRN constantly introduces new and revises existing rules according to new medical knowledge and procedures. However, validating and assigning messages to cancer cases are performed by a \textit{rule engine} of CaReSS.

Below, we provide two example rules based on the examples provided in \cite{OCLRefactor,RCIA}. \textit{Rule 1} is a simple check on the invalid values of two variables, i.e., \textit{basis} and \textit{surgery}. \textit{Rule 2}, on the other hand, is more complex and states that for all the cancer messages of type \textit{H} coming to CaReSS from medical entities (Figure~\ref{overview}), if the \textit{surgery} value is equal to 96 then the \textit{basis} value must be greater than 32. 

\vspace{12pt}
\textbf{Rule 1:}
\begin{equation*}
	basis \neq 71 \land surgery \neq 21
\end{equation*}

\textbf{Rule 2:}
\begin{equation*}
	\forall \; messageType = H \implies (surgery = 96 \implies basis > 32)
\end{equation*}

CaReSS outputs data for researchers to perform epidemiological studies with standard statistical methods and ML techniques.
It also produces cancer statistics used by end-users (e.g., patients and policymakers) as references for well-informed decisions, as shown in Figure~\ref{overview}.

Given that the CaReSS is used to make critical decisions by its stakeholders, it is important that the system is tested thoroughly to ensure its correctness, security, and privacy (e.g., according to the GDPR). Moreover, the data augmentation, transformation, and classification in CaReSS directly affect the quality of the produced data, which are further used to make decisions by the CaReSS's users. Therefore, testing is critical to ensure that the data produced is of the highest quality such that accurate decisions can be made by CaReSS's end-users.   

CRN's team developing the CaReSS is now building automated testing solutions to deal with the constant evolution of the CaReSS, including the changes in the rules it uses, the algorithms it employs, and bugs found. To this end, a set of testing challenges have been identified based on discussion with various people involved in developing the CaReSS. Such challenges are quite common across other healthcare registries; thus, this paper aims to share these challenges together with the investigation of the initial set of proposed solutions.     

We organize the paper as follows: Section~\ref{sec:problem} presents the real-world testing problem we are focusing on, our main objective, and the potential impact that solving the problem will bring. Section~\ref{sec:challenges} explains various challenges, whereas Section~\ref{sec:beyondCRN} will present how these challenges are also valid in other industrial and real-world contexts. Finally, we conclude the paper in Section~\ref{sec:conclusion}.

\section{Scope of the Real-World Problem} \label{sec:problem}

This section presents the real-world problem we focus on for testing in Section~\ref{subsec:problemdetails}, Section~\ref{subsec:objective} explains our key testing objectives, and finally Section~\ref{subsec:expectedimpact} presents the potential impact we would like to achieve with automated testing. 

\subsection{Scope: CaReSS Evolution} \label{subsec:problemdetails}

Automated checking of cancer data against medical rules is an important step in ensuring the quality of patient history registration at CRN and in CaReSS.
However, these medical rules evolve constantly, e.g., because of new findings from medical research and the introduction of new cancer message types from new medical diagnostics or treatments.
Moreover, CaReSS employs algorithms for producing cancer statistics.
These algorithms undergo constant changes because CaReSS keeps upgrading itself by employing advanced versions of algorithms and retraining models when new data or new data types are available.
Such upgrades are necessary to improve the efficiency and timeliness of the cancer data used for medical research and the general public.
In summary, \textit{continuous changes and evolution in data (messages, patients’ histories, rules, etc.) and algorithms lead to the continuous evolution of the CaReSS’s key software components.}

\subsection{Objective: Testing Evolving CaReSS} \label{subsec:objective}
Due to continuous evolution, CaReSS must be continuously tested to guarantee that it produces reliable statistics and data for its end users.
Failing to do so would significantly negatively affect scientific results produced by researchers and potentially mislead patients, hospitals, and policymakers to make inaccurate decisions.
Thus, \textit{a cost-effective, systematic, automated, and self-managing testing approach is needed to handle the evolving aspect of CaReSS!}
More specifically, the innovation planned is a state-of-the-art test infrastructure with cost-effective testing techniques utilizing ML and evolutionary computation algorithms. This will significantly improve the quality of CaReSS and the data and statistics it produces by dealing with the algorithms' continuous evolution and unpredictable behavior. 

\subsection{Expected Impact: Improved CaReSS Quality} \label{subsec:expectedimpact}
First, this innovation is expected to have a long-term impact (of several decades at least) on the service quality that CaReSS provides for its users.
More specifically, the well-tested CaReSS would produce data and statistics with higher quality much more quickly. Such data will be consumed by researchers to perform analyses more accurately and reliably and provide patients, doctors, the general public, and government officials to achieve various objectives.
Second, a systematic and automated testing solution for CaReSS will improve the overall time-wise efficiency of evolving CaReSS for decades, implying a significant reduction in the overall cost of testing and maintaining CaReSS.

\section{Challenges} \label{sec:challenges}

This section presents challenges from three perspectives, i.e., scientific (Section~\ref{subsec:scientific}), project execution (Section~\ref{subsec:projectexecution}), and sharing tools (Section~\ref{subsec:sharing tools}). The scientific challenges are of interest to researchers developing novel testing methods solving these challenges. The project execution and tool sharing challenges are relevant for practitioners who are involved in a similar project as ours in a real-world setting. 

\subsection{Scientific Challenges} \label{subsec:scientific}
New research is needed to propose (modeling) methodologies that can systematically capture various types of the evolution of CaReSS as test models to support automated testing. Examples of evolution include changes in medical rules and software updates. Next, based on the models, test cases can be generated based on test strategies focusing on test objectives, such as finding specific kinds of faults and achieving various types of coverage. To this end, there is a need for new testing strategies, e.g., covering coverage specific to CaReSS, e.g., coverage of rules. Another aspect to investigate is whether classical code coverage (e.g., statement coverage) correlates to rule coverage.

At CRN, an important objective is to test CaReSS in the presence of evolution, which leads to \textit{uncertain behavior} due to \textit{unknown or erroneous data} about a patient’s history, \textit{uncertain time} of the determination of a cancer case, and the uncertain behavior of rules when \textit{emerging or incorrect data} are provided. Solutions for testing software systems under uncertainties (e.g., \cite{zhang2019uncertainty, shin2021uncertainty, camilli2021uncertainty}) have been proposed, which, however, do not address the challenges of testing constantly-evolving software systems like CaReSS, though we acknowledge we can get inspired from the literature on aspects like how to model/specify uncertainties in test models. Existing testing approaches (e.g., \cite{muhlbauer2020identifying, 9678659}) mostly focus on testing evolving software by just looking into their versions. 

The use of ML for decision-making introduces additional uncertainty because of the unpredictable nature of such algorithms. Solutions (e.g., \cite{sun2018testing, 9402064}) have been proposed to test artificial neural networks. Some works (e.g., \cite{catak2022,weiss2022forgotten}) focus on generating highly uncertain test inputs. Such solutions, however, only test ML models in an isolated manner, which is considered insufficient for our context.
We, hence, need \textit{novel test strategies} to test CaReSS in the presence of uncertainty and evolution. \textit{Testing uncertainty and the emerging behaviors of a constantly-evolving healthcare system has not been sufficiently investigated in either the literature or practice. } 

Another area we plan to investigate is uncertainty quantification in the ML models used for decision-making such that we can assess the robustness of these ML models against such uncertainties and possibly reduce such uncertainties. Some works exist that quantify uncertainty in deep learning models~\cite{PURE,NIRVANA,LATTICE}. We want to investigate whether applying such uncertainty quantification methods can generate highly uncertain inputs to test the implemented ML models in isolation and as part of the whole system. Any issues found with such testing would be investigated to see whether these issues can be resolved with additional training data for the ML models or other measures.    

CaReSS is a socio-technical system, i.e., medical coders are an essential part of the system to validate, encode/transform hence creating medical data.
This further complicates the testing of CaReSS.
It is unclear \textit{how to \emph{mock} medical coders (human) in the automated testing without trivializing the system's capabilities}.
One possible direction could be to learn from logs how a human has dealt with different input types and substitute the human with a model during testing.
Although there is some initial research on learning user behavior for testing purposes~\citep{ahlgren:20,salza:22}, these are arguably less complex as the interactions are more trivial than in the medical context where highly-trained domain experts are required.

Due to the dependence on human decisions, CaReSS is currently predominantly tested manually, as previously observed for other systems~\citep{haas:21}.
That is, medical coders use their domain knowledge to manually test CaReSS to ensure it adheres to its (informal) specification.
Due to the complexity of and reliance on medical coders, creating meaningful test cases is complex for software engineers who are not medical experts.
Consequently, supporting medical coders in their manual test efforts, e.g., selecting and prioritizing manual testing as well as manual test quality monitoring~\citep{haas:21}, is an orthogonal (to automated testing) important challenge.

\subsection{Project Execution Challenges} \label{subsec:projectexecution}
\textit{A central challenge concerns the security and privacy of handling sensitive healthcare data.}
Handling patient data is confined to a restricted and secure zone within the premises of CRN.
This includes data access (e.g., for external scientific collaborators with testing expertise), installation of arbitrary software that is needed for testing, and usage of real data for testing purposes.
In particular, the last point is crucial because using real patient data for testing is prohibited by laws and regulations, such as GDPR. 
Alternatives include generating synthetic and realistic data with novel ML approaches (e.g., Generative Adversarial Networks (GANs)). However, this is non-trivial as well, as concerns about the re-identification of real patients from synthetic data and the models trained on real data remain an open issue.
This inhibits or at least complicates sharing of synthetic data generated by as well as the models themselves trained on real patient data.

\subsection{Tool Sharing Challenges} \label{subsec:sharing tools}

\textit{Beyond the data, sharing tools that handle patient data is a challenge on its own.} CaReSS is custom-built and uses a high degree of internally-hosted, proprietary software solutions, which complicates the creation of a \emph{standalone} version for our research purposes. While one option is to rely on open-source systems (if they exist) instead, this decreases the realism of the context in which novel solutions are designed and consequently complicates their adoptions in the real application contexts.

\section{Generalization Beyond the Cancer Registry of Norway} \label{sec:beyondCRN}
The challenges are not specific to CRN, and it can be tailored to serve cancer registries in other countries with reasonable effort, as cancer registration and coding are quite \enquote{universal.}
For instance, ICD-O-3 is commonly referenced for cancer codes.\footnote{\url{https://seer.cancer.gov/icd-o-3/}}
We believe the scientific challenges we identified are commonly faced by cancer registry systems.
In addition, other registries such as the national registry used by tax authorities, election authorities, and other public authorities (e.g., police) face similar challenges.
For instance, tax authorities might automatically execute rules to calculate and validate a person’s or a company’s taxes.

The challenges of re-identification and synthetic user data for testing are also not unique to the CRN case or cancer registries in general.
Other personal, sensitive data require increased protection, such as racial and ethnic origin as well as biometric data, as specified by GDPR.\footnote{\url{https://gdpr-info.eu/recitals/no-51/}}
Software systems handling such data, which should also be tested, can adapt the solutions from the medical domain and the CRN case.

Finally, testing any kind of system where human processing is essential will experience similar challenges related to mocking or learning human behavior, particularly when automated testing is desired.
We believe the solutions proposed for CRN can inform researchers and practitioners working with other systems and in different domains addressing learning domain expert interactions.

\section{Conclusion and Future Work} \label{sec:conclusion}
This paper presented an innovation project taking place at the Cancer Registry of Norway (CRN) to automate the testing of the Cancer Registration Support System (CaReSS) with the ultimate objective of improving the quality of the systems and the data it produces. The data produced by the system are used by key system stakeholders to make key decisions; therefore, ensuring correctness, robustness, and privacy is of utmost importance. We presented the key challenges of automating the testing of CaReSS, including evolution, uncertainty, and dealing with human-in-the-loop. Such automation will revolutionize the current testing practice at CRN, whose impact will result in high-quality CaReSS and the data it produces, thereby positively affecting CaReSS's users. We hope that the presented challenges are interesting for the researchers to investigate since such challenges are common across cancer registries worldwide and other healthcare registries.        

Currently, we are implementing various testing solutions to automate testing CaReSS from various aspects. First, we are looking into automated testing from a human perspective with Graphical User Interface (GUI)-based testing. Second, we are investigating evolutionary algorithms to test CaReSS with its provided REST APIs. Third, we plan to utilize different machine learning algorithms, such as reinforcement learning, to automate testing. Finally, we envision integrating GUI-based and REST API-based testing to develop a more comprehensive testing solution. In addition, we will also develop solutions to handle testing challenges that we described in the paper, including uncertainty and evolution.  

\section{Acknowledgement} \label{sec:acknowledge}
This work is supported by the "AI-Powered Testing Infrastructure for Cancer Registry System" project (Project \#309642) supported by the Research Council of Norway under the Innovation in Public Sector scheme. 

\bibliographystyle{IEEEtran}
\bibliography{IEEEabrv,references.bib}

\begin{thebibliography}{10}
\providecommand{\url}[1]{#1}
\csname url@samestyle\endcsname
\providecommand{\newblock}{\relax}
\providecommand{\bibinfo}[2]{#2}
\providecommand{\BIBentrySTDinterwordspacing}{\spaceskip=0pt\relax}
\providecommand{\BIBentryALTinterwordstretchfactor}{4}
\providecommand{\BIBentryALTinterwordspacing}{\spaceskip=\fontdimen2\font plus
\BIBentryALTinterwordstretchfactor\fontdimen3\font minus
  \fontdimen4\font\relax}
\providecommand{\BIBforeignlanguage}[2]{{%
\expandafter\ifx\csname l@#1\endcsname\relax
\typeout{** WARNING: IEEEtran.bst: No hyphenation pattern has been}%
\typeout{** loaded for the language `#1'. Using the pattern for}%
\typeout{** the default language instead.}%
\else
\language=\csname l@#1\endcsname
\fi
#2}}
\providecommand{\BIBdecl}{\relax}
\BIBdecl

\bibitem{OCLRefactor}
H.~Lu, S.~Wang, T.~Yue, s.~Ali, and J.~F. Nygård, ``Automated refactoring of
  ocl constraints with search,'' \emph{IEEE Transactions on Software
  Engineering}, vol.~45, no.~2, pp. 148--170, 2019.

\bibitem{RCIA}
S.~Wang, T.~Schwitalla, T.~Yue, S.~Ali, and J.~F. Nygård, ``Rcia: Automated
  change impact analysis to facilitate a practical cancer registry system,'' in
  \emph{2017 IEEE International Conference on Software Maintenance and
  Evolution (ICSME)}, 2017, pp. 603--612.

\bibitem{zhang2019uncertainty}
M.~Zhang, S.~Ali, and T.~Yue, ``Uncertainty-wise test case generation and
  minimization for cyber-sphysical systems,'' \emph{Journal of Systems and
  Software}, vol. 153, pp. 1--21, 2019.

\bibitem{shin2021uncertainty}
S.~Y. Shin, K.~Chaouch, S.~Nejati, M.~Sabetzadeh, L.~C. Briand, and F.~Zimmer,
  ``Uncertainty-aware specification and analysis for hardware-in-the-loop
  testing of cyber-physical systems,'' \emph{Journal of Systems and Software},
  vol. 171, p. 110813, 2021.

\bibitem{camilli2021uncertainty}
M.~Camilli, A.~Gargantini, P.~Scandurra, and C.~Trubiani, ``Uncertainty-aware
  exploration in model-based testing,'' in \emph{Proceedings of the 14th {IEEE}
  Conference on Software Testing, Verification and Validation}, ser. {ICST}
  2021.\hskip 1em plus 0.5em minus 0.4em\relax IEEE, 2021, pp. 71--81.

\bibitem{muhlbauer2020identifying}
S.~M{\"u}hlbauer, S.~Apel, and N.~Siegmund, ``Identifying software performance
  changes across variants and versions,'' in \emph{Proceedings of the 35th
  {IEEE}/{ACM} International Conference on Automated Software Engineering},
  ser. {ASE} 2020.\hskip 1em plus 0.5em minus 0.4em\relax {IEEE}, 2020, pp.
  611--622.

\bibitem{9678659}
X.~Wu, ``Effectively analyzing evolving software with differential facts,'' in
  \emph{Proceedings of the 36th {IEEE}/{ACM} International Conference on
  Automated Software Engineering}, 2021, pp. 1064--1068.

\bibitem{sun2018testing}
Y.~Sun, X.~Huang, D.~Kroening, J.~Sharp, M.~Hill, and R.~Ashmore, ``Testing
  deep neural networks,'' \emph{arXiv preprint arXiv:1803.04792}, 2018.

\bibitem{9402064}
Z.~Wang, H.~You, J.~Chen, Y.~Zhang, X.~Dong, and W.~Zhang, ``Prioritizing test
  inputs for deep neural networks via mutation analysis,'' in \emph{Proceedings
  of the 43rd {IEEE}/{ACM} International Conference on Software Engineering},
  ser. {ICSE} 2021, 2021, pp. 397--409.

\bibitem{catak2022}
\BIBentryALTinterwordspacing
F.~O. Catak, T.~Yue, and S.~Ali, ``Uncertainty-aware prediction validator in
  deep learning models for cyber-physical system data,'' \emph{{ACM}
  Transactions on Software Engineering and Methodology}, vol.~31, no.~4, Jul.
  2022. [Online]. Available: \url{https://doi.org/10.1145/3527451}
\BIBentrySTDinterwordspacing

\bibitem{weiss2022forgotten}
M.~Weiss, A.~G. G{\'o}mez, and P.~Tonella, ``A forgotten danger in {DNN}
  supervision testing: Generating and detecting true ambiguity,'' \emph{arXiv
  preprint arXiv:2207.10495}, 2022.

\bibitem{PURE}
\BIBentryALTinterwordspacing
F.~Catak, T.~Yue, and S.~Ali, ``Prediction surface uncertainty quantification
  in object detection models for autonomous driving,'' in \emph{2021 IEEE
  International Conference On Artificial Intelligence Testing (AITest)}.\hskip
  1em plus 0.5em minus 0.4em\relax Los Alamitos, CA, USA: IEEE Computer
  Society, aug 2021, pp. 93--100. [Online]. Available:
  \url{https://doi.ieeecomputersociety.org/10.1109/AITEST52744.2021.00027}
\BIBentrySTDinterwordspacing

\bibitem{NIRVANA}
\BIBentryALTinterwordspacing
F.~O. Catak, T.~Yue, and S.~Ali, ``Uncertainty-aware prediction validator in
  deep learning models for cyber-physical system data,'' \emph{ACM Trans.
  Softw. Eng. Methodol.}, vol.~31, no.~4, jul 2022. [Online]. Available:
  \url{https://doi.org/10.1145/3527451}
\BIBentrySTDinterwordspacing

\bibitem{LATTICE}
\BIBentryALTinterwordspacing
Q.~Xu, S.~Ali, T.~Yue, and M.~Arratibel, ``Uncertainty-aware transfer learning
  to evolve digital twins for industrial elevators,'' in \emph{Proceedings of
  the 30th ACM Joint European Software Engineering Conference and Symposium on
  the Foundations of Software Engineering}, ser. ESEC/FSE 2022.\hskip 1em plus
  0.5em minus 0.4em\relax New York, NY, USA: Association for Computing
  Machinery, 2022, p. 1257–1268. [Online]. Available:
  \url{https://doi.org/10.1145/3540250.3558957}
\BIBentrySTDinterwordspacing

\bibitem{ahlgren:20}
\BIBentryALTinterwordspacing
J.~Ahlgren, M.~E. Berezin, K.~Bojarczuk, E.~Dulskyte, I.~Dvortsova, J.~George,
  N.~Gucevska, M.~Harman, R.~L\"{a}mmel, E.~Meijer, S.~Sapora, and
  J.~Spahr-Summers, ``Wes: Agent-based user interaction simulation on real
  infrastructure,'' in \emph{Proceedings of the 42nd {IEEE}/{ACM} International
  Conference on Software Engineering Workshops}, ser. ICSEW'20.\hskip 1em plus
  0.5em minus 0.4em\relax Association for Computing Machinery, 2020, pp.
  276--284. [Online]. Available: \url{https://doi.org/10.1145/3387940.3392089}
\BIBentrySTDinterwordspacing

\bibitem{salza:22}
\BIBentryALTinterwordspacing
P.~Salza, M.~E. Palma, and H.~C. Gall, ``Synthetic end-user testing: Modeling
  realistic agents based on behavioral examples,'' 2022. [Online]. Available:
  \url{https://arxiv.org/abs/2208.12261}
\BIBentrySTDinterwordspacing

\bibitem{haas:21}
\BIBentryALTinterwordspacing
R.~Haas, D.~Elsner, E.~Juergens, A.~Pretschner, and S.~Apel, ``How can manual
  testing processes be optimized? {D}eveloper survey, optimization guidelines,
  and case studies,'' in \emph{Proceedings of the 29th {ACM} Joint European
  Software Engineering Conference and Symposium on the Foundations of Software
  Engineering}, ser. {ESEC}/{FSE} 2021.\hskip 1em plus 0.5em minus 0.4em\relax
  Association for Computing Machinery ({ACM}), Aug. 2021. [Online]. Available:
  \url{https://doi.org/10.1145/3468264.3473922}
\BIBentrySTDinterwordspacing

\end{thebibliography}

\end{document}